\documentclass[iop]{emulateapj}		
\usepackage{epsfig}			
\usepackage{graphicx,color}		
\usepackage{amssymb}			
\usepackage{color}			
\usepackage{url}			
\usepackage{amsmath}			
\usepackage{rotating}			
\usepackage{float}			
\usepackage{textcomp}			
\usepackage{psfig}
\usepackage{dcolumn}
\usepackage{times}
\usepackage{tabularx}
\usepackage[colorlinks=true,citecolor=blue]{hyperref}
\usepackage[english]{babel}

\shorttitle{Oscillation in coronal bright point}
\shortauthors{T. Samanta et al.}

\begin{document}

\title{Quasi-periodic oscillation of a coronal bright point}

\author{Tanmoy Samanta$^{1}$,
Dipankar Banerjee$^{1}$,
Hui Tian$^{2}$}
 
\affil{$^{1}$Indian Institute of Astrophysics, Koramangala, Bangalore 560034, India. e-mail: {\color{blue}{tsamanta@iiap.res.in}}\\
$^{2}$Harvard-Smithsonian Center for Astrophysics, 60 Garden Street, Cambridge, MA 02138, USA. e-mail: {\color{blue}{hui.tian@cfa.harvard.edu}}}

\begin{abstract}
Coronal bright points (BPs) are small-scale luminous features seen in the solar corona.
Quasi-periodic brightnings are frequently observed in the BPs and are generally linked with underneath magnetic flux changes. 
We study the dynamics of a BP seen in the coronal hole using the
Atmospheric Imaging Assembly (AIA) images,
the Helioseismic and Magnetic Imager (HMI) magnetogram on board the Solar Dynamics Observatory (SDO) 
and spectroscopic data from the newly launched Interface Region Imaging Spectrograph (IRIS). 
The detailed analysis shows that the BP evolves throughout our observing period along with changes in underlying photospheric magnetic flux and shows periodic brightnings in different EUV and FUV images.
With highest possible spectral and spatial resolution of IRIS, we attempted to identify the sources of these oscillations.
IRIS sit-and-stare observation provided a unique opportunity to study the time evolution of one foot point of the BP as the slit position crossed it. 
We noticed enhanced line profile asymmetry, enhanced line width, 
intensity enhancements and large deviation from the average Doppler shift in the line profiles at specific instances which
indicate the presence of sudden flows along the line of sight direction.
We propose that transition region explosive events (EEs) originating from small scale reconnections and the reconnection outflows are affecting the line profiles.
The correlation between all these parameters is consistent with the repetitive reconnection scenario and could explain the quasi-periodic nature of the brightening.
\end{abstract}


\section{Introduction}
Coronal bright points (BPs) are bright dynamical features seen in quiet-sun and coronal holes. 
The dynamics and evolution was studied in X-rays and EUV wavelengths 
\citep{1973ApJ...185L..47V,1974ApJ...189L..93G,1976SoPh...49...79G,1976SoPh...50..311G,1981SoPh...69...77H,2001SoPh..198..347Z,2008ApJ...681L.121T,2012ApJ...746...19Z,2013NewA...23...19L}.
They generally live for a few hours to few days and have sizes less than $50''$.
BPs are believed to be composed of loops connected locally with the photospheric bipolar magnetic fields \citep{1976SoPh...50..311G,1979SoPh...63..119S}. 
With recent high resolution EUV images it is clear that a BP is not a point or simple loop-like structure
but looks like a miniature active region with multiple magnetic poles with several connectivities.  
Moreover, depending on the emergence and cancellation of the magnetic polarities, the BPs evolve with time and show a lot of dynamics. 
Theoretical model argues that the interaction between two opposite polarities creates a X-point  magnetic reconnection that
locally heats the corona and produces BPs \citep{1994ApJ...427..459P,1994SoPh..151...57P}. 
The locations of BPs appear to be related to the giant convection cells \citep{2014ApJ...784L..32M}.
\cite{2012ApJ...746...19Z,2014A&A...568A..30Z} suggested that small bipolar emerging magnetic loops might reconnect with an overlying large loop
or open field lines and produce brightenings. They also proposed that BPs might consist of two components,
one is long-lived smooth component due to gentle quasi-separatrix layer (QSL) reconnections and other is quasi-periodic
impulsive component, called as BP flashes.

\cite{1979SoPh...63..119S} reported that the BPs  evolve with a  6 minutes time scale.
Several observations in X-ray and EUV show periodic variation in the intensity of BPs over a broad range of periodicity 
\citep{1979SoPh...63..119S,1979SoPh...63..113N,1981SoPh...69...77H,1992PASJ...44L.161S,2011MNRAS.415.1419K,2008A&A...489..741T,2011A&A...526A..78K,2013SoPh..286..125C}.
Some suggested that these oscillations are caused by the leakage of acoustic waves (p-modes), 
which propagate along the magnetic flux tubes and converts into magnetoacoustic mode at higher atmosphere
\citep{2003ApJ...599..626B, 2008AnGeo..26.2983K, 2010MNRAS.405.2317S}. 
Others believe that the intensity oscillations are  due to repeated magnetic reconnections. 
\citep{2003A&A...398..775M,2004A&A...418..313U, 2006A&A...446..327D}.

Several studies have been carried out to understand the periodic nature, but their origin
remains inconclusive. Here, we study a BP inside a coronal hole (CH) as seen in 
the AIA EUV coronal images \citep{2012SoPh..275...17L} and in HMI magnetogram \citep{2012SoPh..275..229S} on the SDO
and simultaneously with the newly launched IRIS \citep{2014SoPh..289.2733D}.
Combining imaging and spectroscopic observations, we study the dynamical changes within this bright point and  its variability. 
We show that the time variability can be explained in terms of a repeated magnetic reconnection scenario.

\section{Data analysis and Results}

\subsection{Observation and Data Reduction}
\begin{figure*}
\centering
\includegraphics[angle=90,trim = -35mm 0mm 0mm -1mm,clip,width=5.0cm]{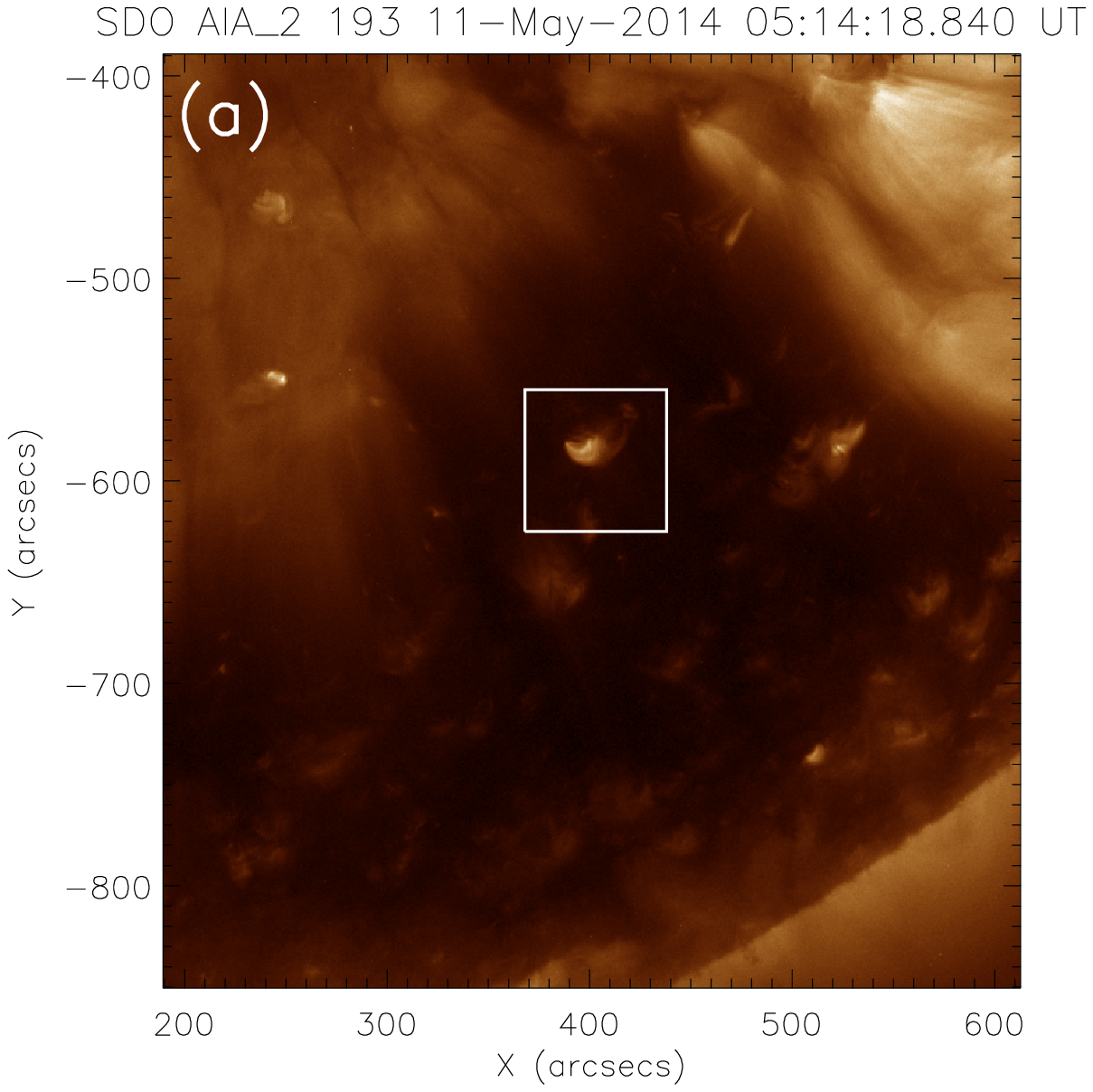}\includegraphics[angle=90, clip,width=13.0cm]{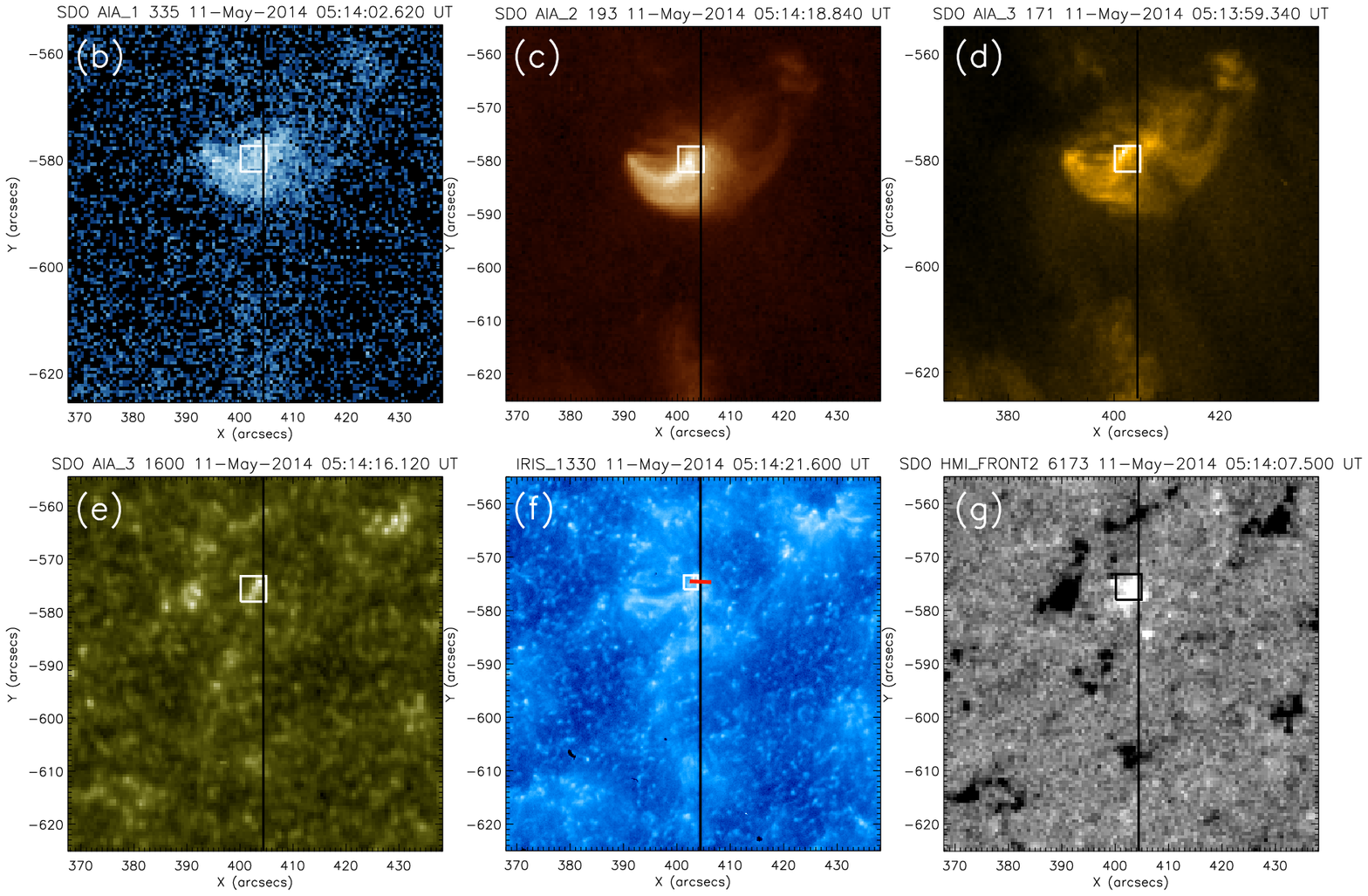}
\caption{\textrm{(a)}: AIA 193 \r{A} image showing a coronal hole. 
The white box represents our Region Of Interest (ROI), 
covering a bright point in the coronal hole.
Zoomed view of the ROI as recorded by different AIA channels are shown in \textrm{(b-e)}, 
\textrm{(f)} shows IRIS 1330 \r{A} SJI image and \textrm{(g)} shows HMI line of sight (LOS) magnetogram.  
The vertical black line on each image represents the position of the IRIS slit.
Intensity within small white box of \textrm{(b-f)} are used to study oscillation properties (see Figure~\ref{aia_lc}).
The red horizontal tick marked on IRIS 1330 \r{A} image \textrm{(f)} is the location 
where we study the variation of different line parameters of the \ion{Si}{4} 1393.76 \r{A} line (see Figure~\ref{si_lc}). A movie of the ROI is available online (Movie~1).}
\label{aia_iris_hmi} 
\end{figure*}

Observational data was obtained from IRIS, AIA and
HMI instruments form 5:14~UT to 6:34~UT on 11 May 2014. 
We used AIA images centered at
335~\r{A}, 193~\r{A}, 171~\r{A} and 1600~\r{A}. 
AIA and HMI images have $0.6''$ pixel size and were co-aligned.
IRIS data was taken in sit-and-stare mode.
It was pointing towards a coronal hole (centered at $404'', -612''$).
Slit-jaw images (SJI) were available only with the 1330~\r{A} filter. 
We have used  IRIS Level~2 processed data which takes care of
dark current, flat field and geometrical corrections etc.
The exposure time and cadence of 1330~\r{A} SJI and spectra were 8~sec and 9.6~sec respectively. 
IRIS have a pixel size of $0.166''$.
AIA and HMI data was then co-aligned with IRIS data. 
IRIS 1330~\r{A} and AIA 1600~\r{A} were used for co-alignment.
De-rotation was performed on all AIA and HMI images to co-align the data cubes.

Figure~\ref{aia_iris_hmi}~(a) shows the coronal hole in  AIA 193~\r{A} image. 
The magnetogram (Figure~\ref{aia_iris_hmi}~(g)) shows that the CH is dominated by negative polarity magnetic field (colorbar can be seen in Figure~\ref{hmi}). 
We made a 48~hours movie with AIA 193~\r{A} images and HMI magnetograms with 1~hour cadence.
It reveals that the BP appears around 16:00~UT on 10~May~2014  with emergence of some positive flux 
and disappears around 17:00~UT on 12~May~2014 (with a life time $\sim$~38~hours) with the complete disappearance of the positive flux.
The positive flux cocentration within this BP over the dominated background negative flux is the probable reason of the existence of the BP.
Figure~\ref{aia_iris_hmi} shows the BP as seen in various AIA channels \textrm{(b-e)},
IRIS 1330~\r{A} SJI image (f) and HMI LOS magnetogram (g).  
The vertical black line on each image represents the position of the IRIS slit.
It clearly shows that the IRIS slit is crossing one foot point of the bright point, where the magnetic field is positive.

\subsection{Magnetic field evolution}

\begin{figure}
\centering
\includegraphics[angle=90,width=8.6cm]{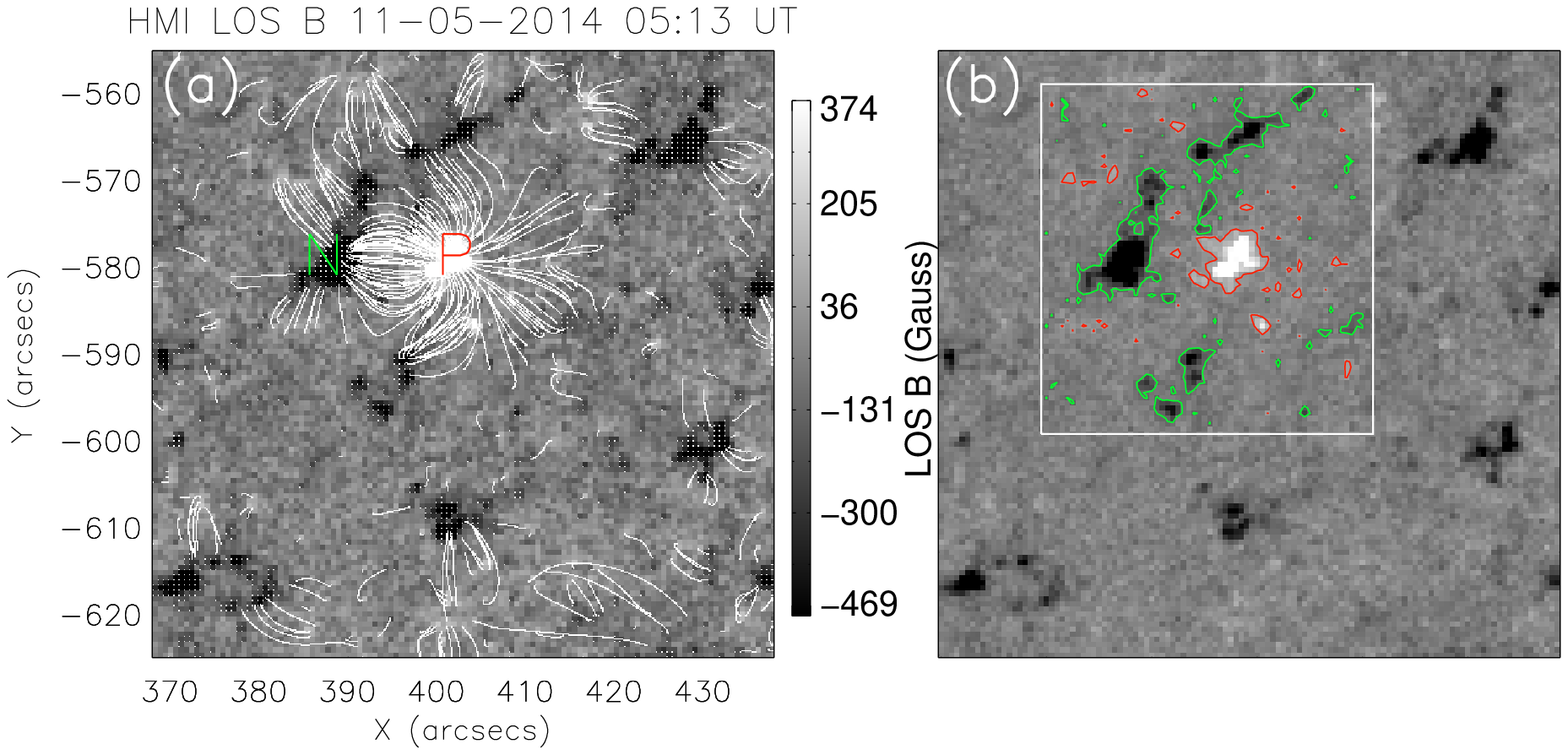}
\includegraphics[angle=90,width=8.6cm]{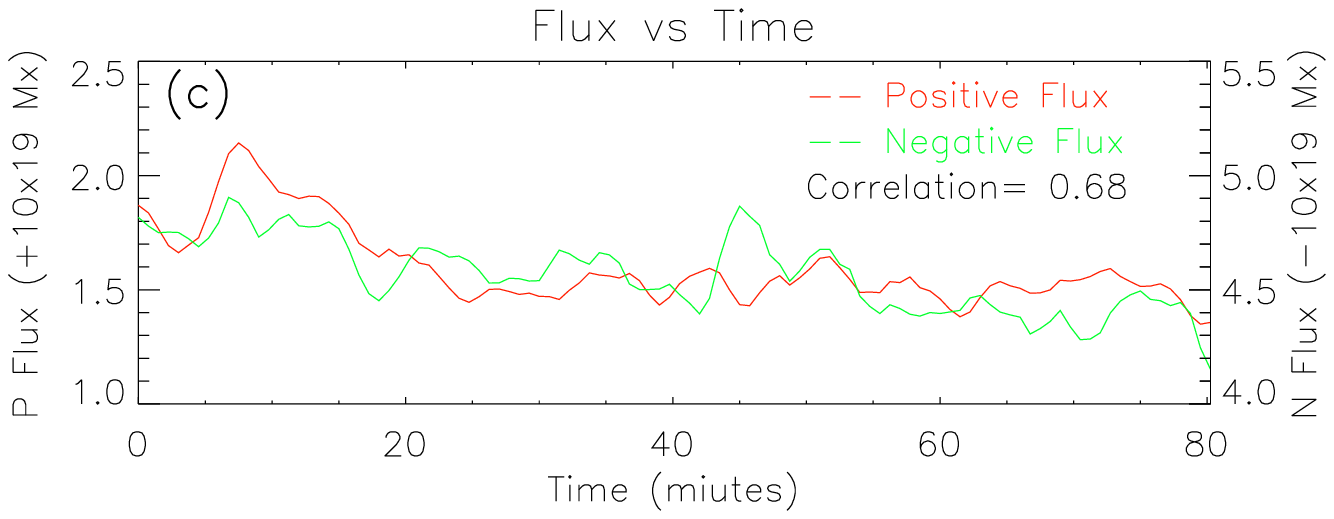}
\caption{\textrm{(a)}: HMI LOS magnetogram and extrapolated potential field lines (white lines). 
\textrm{(b)}: On the same magnetogram the red and green contours represent the field above +20 and -20~Gauss respectively. 
The magnetic fluxes (positive and negative) are calculated within the contours (red and green respectively) inside the rectangular white box.
\textrm{(c)}: The variation of measured positive and negative flux over time.}
\label{hmi} 
\end{figure}
Figure~\ref{hmi}~(a) shows the HMI line of sight (LOS) magnetogram. 
Potential field extrapolation was performed by assuming a constant $\alpha$ force-free magnetic field (with
$ \alpha = 0)$ \citep{1972SoPh...25..127N, 1981A&A...100..197A}.
The white lines connecting different polarities ($ \ge \pm 20$ Gauss) represent the extrapolated field lines. 
Strong connectivity between primary polarity P (positive) and N (negative) can be seen.
It agrees well with the intensity images as seen in IRIS SJI and AIA images. 
Our ROI lies well within a CH, primarily dominated by negative flux over the entire region.
Based on the main connectivity with the positive flux (P), we have selected a region marked as a white box  in Figure~\ref{hmi}~(b) 
to calculate the magnetic flux.
Contours with +20 and -20 Gauss are drawn in red and green respectively and fluxes are calculated within these contours.
After calculating the fluxes for each time frame, smoothed over 3 frames, light curves (LC) are shown in Figure~\ref{hmi}~(c).
There appears to be a good correlation between positive and negative flux with a correlation coefficient (CC) of 0.68. 
It indicates that both positive and negative flux amplitudes vary in a similar manner.
Movie~1 (available online) shows small scale magnetic flux emergence and cancellation. 
It is possible that  newly emerging small bipolar 
loops are reconnecting with the overlying pre-existing large loop (P-N) as suggested by \citep{1994ApJ...427..459P,2012ApJ...746...19Z,2014A&A...568A..30Z}, which could explain the good correlation.
 In section \ref{sp},  we show signatures of reconnection and their relation with flux changes.

\subsection{Imaging observations}

\begin{figure}
\centering
\includegraphics[angle=90,clip,width=4.55cm]{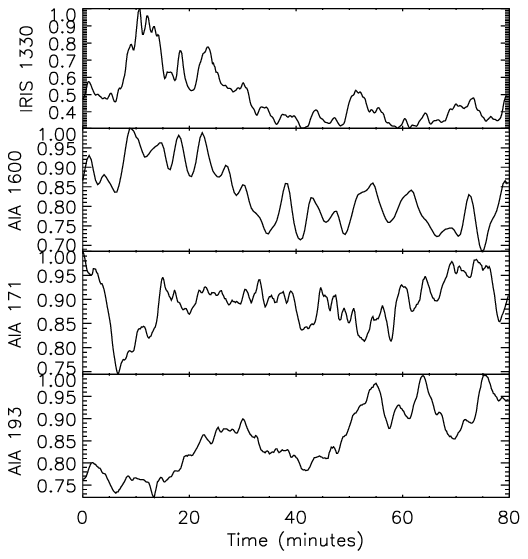}\includegraphics[angle=90,trim = -0.35mm 0mm 0mm 0mm,clip,width=2.045cm]{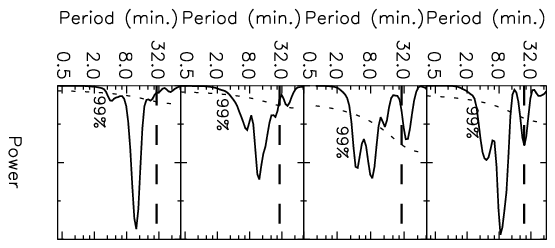}\includegraphics[angle=90,trim = -5.5mm -1.1mm -5mm -1mm,clip,width=2.03cm]{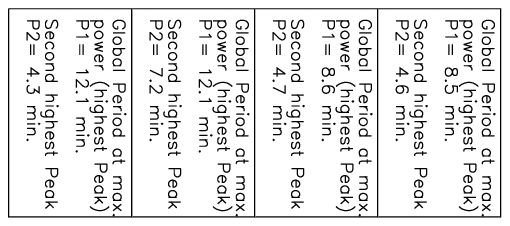}
\caption{ In each row the left panel shows the variation of intensities of different AIA and IRIS channel. 
Middle panel shows the global wavelet power spectrum. The confidence levels are overplotted with dashed lines. 
The right panels show the significant periods as measured from the global wavelet spectrum. 
Intensity variations of AIA images and IRIS 1330~\r{A} SJI images are calculated from the region inside the white boxes as shown in the Figure~\ref{aia_iris_hmi}~\textrm{(b-f)}.}
\label{aia_lc} 
\end{figure}

We have focused on the dynamics of one foot point (positive polarity P). 
We study the time evolution of this small region as seen with simultaneous multi-wavelength images corresponding to the transition region (TR) and coronal layers. 
We have computed average intensities inside the small white  box on top of the foot-point as shown in Figure~\ref{aia_iris_hmi}~(b-f).
The size of the white box in AIA was 8x8 pixel ($ \sim (4.8 '')^2$) and IRIS was 15x15 pixel ($ \sim (2.5 '')^2$).
AIA box size  was selected slightly larger to accommodate the loop expansion higher up and also to reduce movement effects of the loop if any. 
The smoothed light curves (LCs) over 3 time frames are shown in the Figure~\ref{aia_lc}. 
Left panels show (from top to bottom) the LCs of IRIS 1330~\r{A}, AIA 1600~\r{A}, 171~\r{A} and 193~\r{A} respectively.
We have performed wavelet analysis \citep{1998BAMS...79...61T} on each LC after removing background trend. 
We use Morlet function, a complex sine wave modulated by a Gaussian, for convolution with the time series in the wavelet transform.
The global wavelet power spectrum are shown in the middle panel of Figure \ref{aia_lc}. A confidence level of 99\% is overplotted by dotted white line. 
The confidence level was set by assuming white noise \citep{1998BAMS...79...61T}.
Measured periods are printed on the right panels.
The global wavelet power plots clearly show the presence of periodicities.
A dominant period around 8 minutes is present in IRIS 1330~\r{A} and AIA 1600, 171~\r{A} LCs. 
Though  the global wavelet plots of  AIA 171~\r{A} and 193~\r{A} show the presence of strongest peak $\sim$12 minutes there is a weaker power around 8 minutes.

Now we try to explore the source of this oscillations. 
Several observational evidence  show a positive correlation between the EUV and
X-ray emission with the underlying photospheric magnetic flux \citep{1999ApJ...510L..73P, 2001ApJ...547.1100H, 2004A&A...418..313U, 2008A&A...492..575P}. 
\cite{1999ApJ...510L..73P} have observed that X-ray and EUV emissions are  temporally 
correlated with the photospheric magnetic flux. They also suggested the possibility of several small ``network flares'' occurring during the lifetime of these BPs.
\cite{2013SoPh..286..125C} have found a good temporal correlation between magnetic flux
associated with the foot points and the intensity brightening and suggested that the possibility of repeated reconnection scenario.
Note that most of these were imaging observations. In the following subsection we study the time evolution from the  IRIS spectroscopic data 
which provides additional information of the possible sources of these oscillations.
IRIS sit-and-stare observation provides an ideal opportunity to study the time evolution of this foot point as the slit position is crossing it. 
\subsection{Spectroscopic Analysis}
\label{sp}

At first, a single Gaussian fit was performed on the averaged (over all the pixel along the slit and time) profile of photospheric \ion{Si}{1} 1401.513~\r{A} line
for absolute calibration of wavelength.
Now, for our spectroscopic study, we  have selected a position ($402.40'', -574.52''$) which corresponds to one foot point of the bright point loop system.
The position is marked by a red tick mark on  IRIS 1330~\r{A} SJI in Figure~\ref{aia_iris_hmi}~(f). 
An average over 3 pixels along the slit and a running average of 3 points along dispersion was applied
to the spectra to improve signal-to-noise ratio.
After that, a single Gaussian fit was applied to each IRIS \ion{Si}{4} 1393.76~\r{A} line profile to derive line intensity,
Doppler shift, and FWHM of the line. To compute the asymmetry in the line profile, 
we performed Red-Blue (RB) asymmetry analysis \citep{2009ApJ...701L...1D,2011ApJ...738...18T,2011ApJ...732...84M}.
A single Gaussian fit was performed only in the core of the profile to find line centroid (similar to that in \citet{2014Sci...346A.315T}).
Red and blue wings was then subtracted and normalized to peak intensity to construct RB asymmetry profile (in percentage).
Finally, we have constructed the RB asymmetry light curve by taking average over 15 - 40 Km/s velocity range for each profile.
Positive and negative values represent enhancements in the red and blue wings respectively.

\begin{figure}
\centering
\includegraphics[angle=90,clip,width=5.02cm]{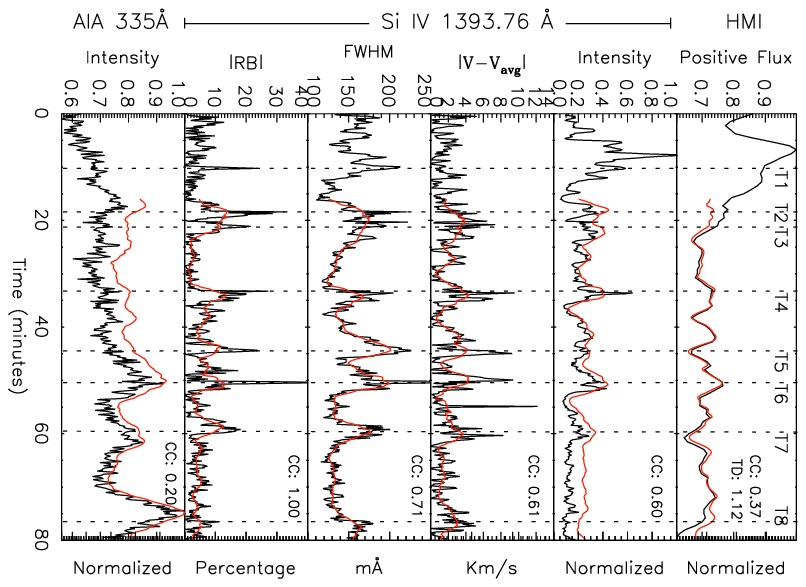}\includegraphics[angle=90,trim = -1.3mm 0mm 0mm 0mm,clip,width=1.78cm]{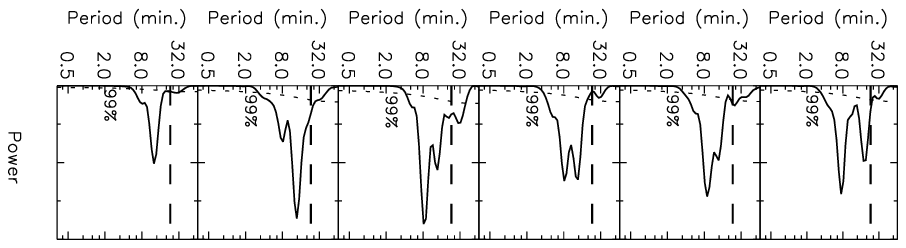}\includegraphics[angle=90,trim = -6.2mm -1.1mm -5mm -1mm,clip,width=1.78cm]{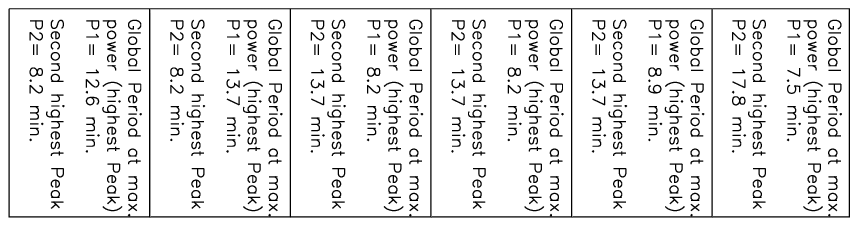}
\caption{Similar to Figure~\ref{aia_lc}, top row corresponds to HMI positive flux and bottom row corresponds to AIA 335~\r{A} intensity 
and other rows correspond to different line parameters as calculated from the \ion{Si}{4} 1393.76~\r{A} line profiles. 
$|RB|$ and $|V-V_{avg}|$ represents absolute RB asymmetry and the deviation from average Doppler shift respectively, where $V_{avg}$ is the average Doppler shift.
Red lines are the trend subtracted smoothed curves for duration 16 to 80 minutes.
\ion{Si}{4} 1393.76~\r{A} line profiles correspond to the location as marked in Figure~\ref{aia_iris_hmi}~(f) with the red tick mark on the IRIS 1330~\r{A} SJI image. 
The correlation coefficient (CC) between all the LCs with the absolute RB is printed on the respective panel.}
\label{si_lc} 
\end{figure}

\begin{figure}
\centering
\includegraphics[angle=90,clip,width=8.4cm]{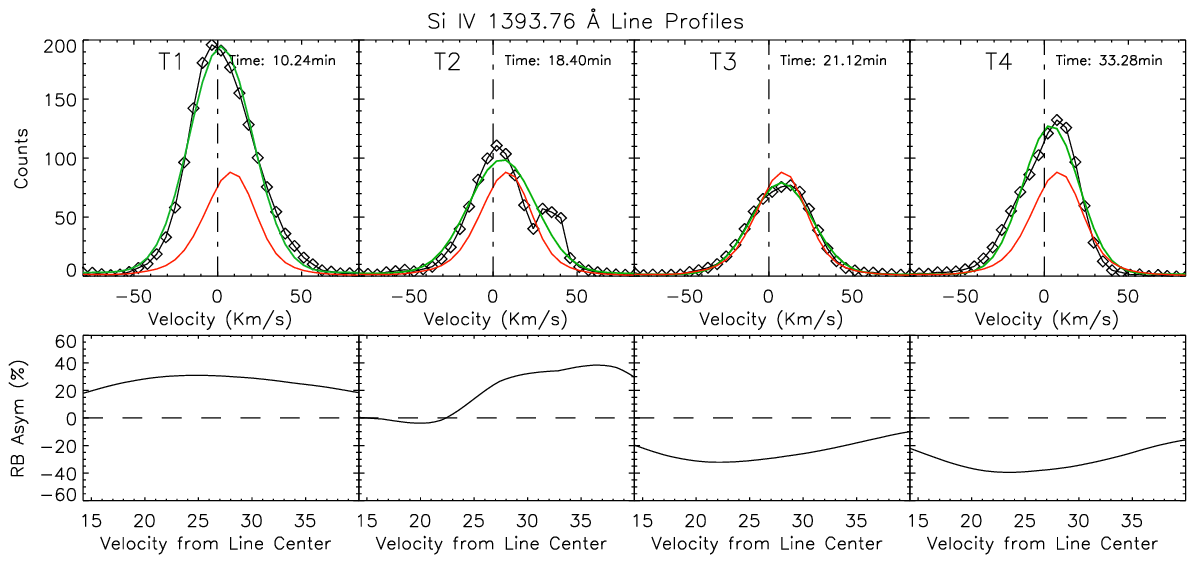}
\includegraphics[angle=90,clip,width=8.4cm]{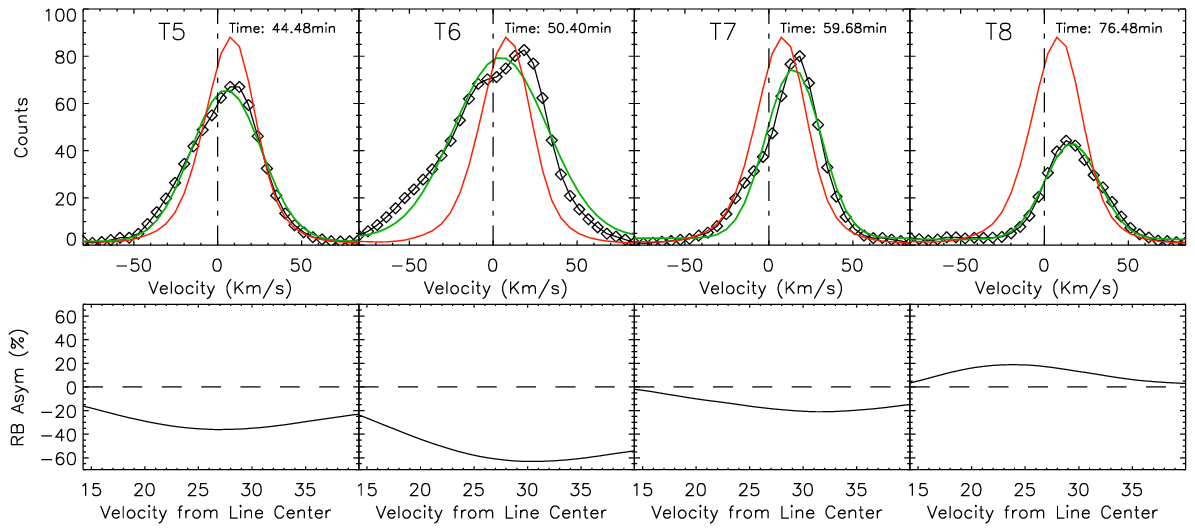}
\caption{The panels show line profiles of the \ion{Si}{4} 1393.76~\r{A} line at various instances as labeled along with their corresponding RB asymmetry profile (from 15 to 40 Km/s) in the bottom. 
Red lines represent average line profile over time. Green lines are the
Gaussian fits. A movie (Movie~2) of the line profiles is available online. }
\label{line_profile} 
\end{figure}

Now, the variation of all the line parameters with time along with HMI positive flux and AIA 335~\r{A} intensity are shown in Figure~\ref{si_lc}. 
Different rows (from top to bottom) correspond to  HMI positive flux, intensity, 
deviation from average Doppler shift ($|V-V_{avg}|$), FWHM, absolute RB asymmetry ($|RB|$) of \ion{Si}{4} 1393.76~\r{A} line and AIA 335~\r{A} intensity respectively. 
Now for the correlation studies between different parameters and to find out the periodic nature, we have selected a time interval of 16-80 minutes. 
The initial rapid change in flux affects the power analysis, so we have omitted first 16 minutes for this analysis.
We use trend subtracted smoothed LCs as represented by red lines in Figure~\ref{si_lc} for easier comparison.
The wavelet analysis was applied over the smoothed curve to investigate the oscillation properties.
Global wavelet power spectrum along with 99\% confidence level are shown in middle panels.
Observed periodicities are then printed on the right side.
The power analysis on the line parameters show the omnipresence of a strong periodicity around 8 and 13 minutes. 
Correlation coefficients (CC) of all the line parameters with the absolute RB asymmetry 
are printed on each panel. We correlate different LCs with RB asymmetry LC as the RB asymmetry provides a good measurement of the distortion in line profile form Gaussian. 
To investigate it further, we will focus on specific instances and will have a closer look at the variation of line profiles.
We notice sudden changes in the line parameters at certain times. 
In Figure~\ref{si_lc}, horizontal grids represent the particular instances where certain changes occur.
The line profiles at T1, T2, T3, T4, T5, T6, T7 and T8 time instances are shown in 
Figure~\ref{line_profile} along with averaged (in red) and Gaussian fit (in green) profile.
The RB asymmetry profiles are shown in the lower panel. 
It clearly shows strong asymmetry in the profiles at those particular instances due to existence of some secondary emission component. 
Movie~2 (available online) shows the evolution of the line profiles.

The correlation between intensity, FWHM, $|V-V_{avg}|$ and $|RB|$ can be easily explained in terms of sudden flows along the line of sight. 
In active region boundaries, all the line parameters coherently changes due to quasi-periodic upflows in the medium \citep{2010ApJ...722.1013D, 2011ApJ...727L..37T,2012ApJ...759..144T}.
In our study, we observed both red and blueward asymmetries time to time (see Figure~\ref{line_profile}) which is different from AR boundary study where predominantly blueward asymmetry is reported.
Our observations can be  explained in terms of  both high-speed upflows and downflows. 
Transition region explosive events (EEs) could be a possible explanation \citep{1983ApJ...272..329B, 1989ApJ...345L..95D, 1997Natur.386..811I, 1998ApJ...497L.109C, 2004A&A...427.1065T,2014ApJ...797...88H}.
EEs are believed to result from reconnection which will affect the line profiles.
Reconnection jets (regardless of the direction) usually lead to enhancements at the line wing (or wings), 
which would lead to large RB values, larger line width, larger intensity. 
The presence of these additional flow components usually also leads to large perturbation of the Doppler shift.
This sudden changes in the line parameters is consistent with the behavior of bursty EEs. 
We find periodic occurrence of EEs at one footpoint of a coronal BP. 
The observed periodicities at one footpoint of the BP at the different 
temperature channels is similar to the observed periodic changes in different line parameters which may indicate that recurring EEs are likely producing the oscillatory signal seen in the BP.

\cite{1998ApJ...497L.109C} reported that EEs happen preferentially in regions with weak and mixed polarity magnetic fluxes.
They also noticed that majority of EEs occur during the cancellation of photospheric magnetic field. 
In our study, we find correlation between RB asymmetry and underneath positive flux. 
We use cross correlation technique to find if the two LCs are showing similar variation with some time delay (similar to \citet{2012ApJ...759..144T}).
It shows that $|RB|$ LC have a time delay (TD) of 1.12 minutes from magnetic flux LC.
Hence, It can be conjectured that the line profiles are 
strongly affected during magnetic flux cancellation phase. It could be that the new emerging flux reconnects with pre-existing field which cancels the local flux and creates EEs. 
The periodic behavior might be explained by repeated reconnection.
One can also notice while carefully looking at the variations, that the RB asymmetry do not change randomly, 
it changes slowly with a sharp increase around specific instances and then slowly decreases. 
This matches well with a slow reconnection scenario \citep{1993SoPh..143..119W}. Slow magnetic reconnections occurring in lower atmosphere could initiate fast reconnection in the TR and then decreases slowly. 
After some relaxation time it repeats. 

We also searched for the signature of heating due to reconnections. The AIA 335\r{A} channel corresponds to ionization temperature of about 2.5~x~$10^6 $ K with a wide temperature response. In Figure~\ref{si_lc}, the 335~\r{A} light curve 
shows increase of intensity during the time when the line profiles show sudden changes. Due to reconnection, magnetic energy releases and heats the medium locally at TR.
This localized heating close to the reconnection point may increase temperature at certain pockets in the TR and AIA may see some of these emissions.
Hence, depending on the heating, higher temperature emission can enhance
during the reconnection time. The AIA 335~\r{A} light curves show similar property. We wish to address this conjecture in our future work while looking at coronal spectral lines.

\section{Conclusion}
We study the dynamics of a bright point within a coronal hole using combined imaging, spectroscopic and magnetic measurements. 
We focused our analysis on one foot point of a bipolar loop structure. 
Throughout our observation, both positive and negative magnetic flux shows correlated variations.
This may suggest that  emerging flux interacts with the pre-existing overlying fields, 
which results in reconnection and cancellation of the flux at the site of bright point.
We conjecture that the periodic behavior of the positive flux may correspond to repeated reconnections
which leads to a series of continuous periodic brightenings as seen in different EUV and FUV lines. 
We propose that EEs are created due to X-point magnetic reconnection and resultant outflows generally affect the line profiles.
The presence of the secondary component  emission in the line profiles confirms that.
Furthermore, we observed enhanced line profile asymmetry, enhanced line width, large deviation from the average Doppler shift at specific instances. 
The correlation between all these parameters is consistent with the scenario of repetitive alteration of the line profile by bursty reconnection outflows. 
We not only observe similar periodicities at different line parameters of \ion{Si}{4} line but 
also with AIA channels and they are concurrent in time and space and hence seems to be related. During EEs (see Fig.~\ref{si_lc})
we see a corresponding change in the magnetic field - this is a one-to-one correspondence and certainly indicates the close relationship between the two.

\acknowledgments
We thank the IRIS team for proving the data in the public domain. 
IRIS is a NASA small explorer mission developed and operated by 
LMSAL with mission operations executed at NASA Ames Research center 
and major contributions to downlink communications funded by the Norwegian Space Center (NSC, Norway) through an ESA PRODEX contract.
H. T. is supported by contracts 8100002705 and SP02H1701R from Lockheed-Martin to SAO.


\begin{thebibliography}{49}
\expandafter\ifx\csname natexlab\endcsname\relax\def\natexlab#1{#1}\fi

\bibitem[{{Alissandrakis}(1981)}]{1981A&A...100..197A}
{Alissandrakis}, C.~E. 1981, \aap, 100, 197

\bibitem[{{Bogdan} {et~al.}(2003){Bogdan}, {Carlsson}, {Hansteen}, {McMurry},
  {Rosenthal}, {Johnson}, {Petty-Powell}, {Zita}, {Stein}, {McIntosh}, \&
  {Nordlund}}]{2003ApJ...599..626B}
{Bogdan}, T.~J., {et~al.} 2003, \apj, 599, 626

\bibitem[{{Brueckner} \& {Bartoe}(1983)}]{1983ApJ...272..329B}
{Brueckner}, G.~E., \& {Bartoe}, J.-D.~F. 1983, \apj, 272, 329

\bibitem[{{Chae} {et~al.}(1998){Chae}, {Wang}, {Lee}, {Goode}, \&
  {Sch{\"u}hle}}]{1998ApJ...497L.109C}
{Chae}, J., {Wang}, H., {Lee}, C.-Y., {Goode}, P.~R., \& {Sch{\"u}hle}, U.
  1998, \apjl, 497, L109

\bibitem[{{Chandrashekhar} {et~al.}(2013){Chandrashekhar}, {Krishna Prasad},
  {Banerjee}, {Ravindra}, \& {Seaton}}]{2013SoPh..286..125C}
{Chandrashekhar}, K., {Krishna Prasad}, S., {Banerjee}, D., {Ravindra}, B., \&
  {Seaton}, D.~B. 2013, \solphys, 286, 125

\bibitem[{{De Pontieu} \& {McIntosh}(2010)}]{2010ApJ...722.1013D}
{De Pontieu}, B., \& {McIntosh}, S.~W. 2010, \apj, 722, 1013

\bibitem[{{De Pontieu} {et~al.}(2009){De Pontieu}, {McIntosh}, {Hansteen}, \&
  {Schrijver}}]{2009ApJ...701L...1D}
{De Pontieu}, B., {McIntosh}, S.~W., {Hansteen}, V.~H., \& {Schrijver}, C.~J.
  2009, \apjl, 701, L1

\bibitem[{{De Pontieu} {et~al.}(2014){De Pontieu}, {Title}, {Lemen}, {Kushner},
  {Akin}, {Allard}, {Berger}, {Boerner}, {Cheung}, {Chou}, {Drake}, {Duncan},
  {Freeland}, {Heyman}, {Hoffman}, {Hurlburt}, {Lindgren}, {Mathur}, {Rehse},
  {Sabolish}, {Seguin}, {Schrijver}, {Tarbell}, {W{\"u}lser}, {Wolfson},
  {Yanari}, {Mudge}, {Nguyen-Phuc}, {Timmons}, {van Bezooijen}, {Weingrod},
  {Brookner}, {Butcher}, {Dougherty}, {Eder}, {Knagenhjelm}, {Larsen},
  {Mansir}, {Phan}, {Boyle}, {Cheimets}, {DeLuca}, {Golub}, {Gates}, {Hertz},
  {McKillop}, {Park}, {Perry}, {Podgorski}, {Reeves}, {Saar}, {Testa}, {Tian},
  {Weber}, {Dunn}, {Eccles}, {Jaeggli}, {Kankelborg}, {Mashburn}, {Pust},
  {Springer}, {Carvalho}, {Kleint}, {Marmie}, {Mazmanian}, {Pereira}, {Sawyer},
  {Strong}, {Worden}, {Carlsson}, {Hansteen}, {Leenaarts}, {Wiesmann},
  {Aloise}, {Chu}, {Bush}, {Scherrer}, {Brekke}, {Martinez-Sykora}, {Lites},
  {McIntosh}, {Uitenbroek}, {Okamoto}, {Gummin}, {Auker}, {Jerram}, {Pool}, \&
  {Waltham}}]{2014SoPh..289.2733D}
{De Pontieu}, B., {et~al.} 2014, \solphys, 289, 2733

\bibitem[{{Dere} {et~al.}(1989){Dere}, {Bartoe}, {Brueckner}, \&
  {Recely}}]{1989ApJ...345L..95D}
{Dere}, K.~P., {Bartoe}, J.-D.~F., {Brueckner}, G.~E., \& {Recely}, F. 1989,
  \apjl, 345, L95

\bibitem[{{Doyle} {et~al.}(2006){Doyle}, {Popescu}, \&
  {Taroyan}}]{2006A&A...446..327D}
{Doyle}, J.~G., {Popescu}, M.~D., \& {Taroyan}, Y. 2006, \aap, 446, 327

\bibitem[{{Golub} {et~al.}(1974){Golub}, {Krieger}, {Silk}, {Timothy}, \&
  {Vaiana}}]{1974ApJ...189L..93G}
{Golub}, L., {Krieger}, A.~S., {Silk}, J.~K., {Timothy}, A.~F., \& {Vaiana},
  G.~S. 1974, \apjl, 189, L93

\bibitem[{{Golub} {et~al.}(1976{\natexlab{a}}){Golub}, {Krieger}, \&
  {Vaiana}}]{1976SoPh...49...79G}
{Golub}, L., {Krieger}, A.~S., \& {Vaiana}, G.~S. 1976{\natexlab{a}}, \solphys,
  49, 79

\bibitem[{{Golub} {et~al.}(1976{\natexlab{b}}){Golub}, {Krieger}, \&
  {Vaiana}}]{1976SoPh...50..311G}
---. 1976{\natexlab{b}}, \solphys, 50, 311

\bibitem[{{Habbal} \& {Withbroe}(1981)}]{1981SoPh...69...77H}
{Habbal}, S.~R., \& {Withbroe}, G.~L. 1981, \solphys, 69, 77

\bibitem[{{Handy} \& {Schrijver}(2001)}]{2001ApJ...547.1100H}
{Handy}, B.~N., \& {Schrijver}, C.~J. 2001, \apj, 547, 1100

\bibitem[{{Huang} {et~al.}(2014){Huang}, {Madjarska}, {Xia}, {Doyle},
  {Galsgaard}, \& {Fu}}]{2014ApJ...797...88H}
{Huang}, Z., {Madjarska}, M.~S., {Xia}, L., {Doyle}, J.~G., {Galsgaard}, K., \&
  {Fu}, H. 2014, \apj, 797, 88

\bibitem[{{Innes} {et~al.}(1997){Innes}, {Inhester}, {Axford}, \&
  {Wilhelm}}]{1997Natur.386..811I}
{Innes}, D.~E., {Inhester}, B., {Axford}, W.~I., \& {Wilhelm}, K. 1997, \nat,
  386, 811

\bibitem[{{Kariyappa} {et~al.}(2011){Kariyappa}, {Deluca}, {Saar}, {Golub},
  {Dam{\'e}}, {Pevtsov}, \& {Varghese}}]{2011A&A...526A..78K}
{Kariyappa}, R., {Deluca}, E.~E., {Saar}, S.~H., {Golub}, L., {Dam{\'e}}, L.,
  {Pevtsov}, A.~A., \& {Varghese}, B.~A. 2011, \aap, 526, A78

\bibitem[{{Kumar} {et~al.}(2011){Kumar}, {Srivastava}, \&
  {Dwivedi}}]{2011MNRAS.415.1419K}
{Kumar}, M., {Srivastava}, A.~K., \& {Dwivedi}, B.~N. 2011, \mnras, 415, 1419

\bibitem[{{Kuridze} {et~al.}(2008){Kuridze}, {Zaqarashvili}, {Shergelashvili},
  \& {Poedts}}]{2008AnGeo..26.2983K}
{Kuridze}, D., {Zaqarashvili}, T.~V., {Shergelashvili}, B.~M., \& {Poedts}, S.
  2008, Annales Geophysicae, 26, 2983

\bibitem[{{Lemen} {et~al.}(2012){Lemen}, {Title}, {Akin}, {Boerner}, {Chou},
  {Drake}, {Duncan}, {Edwards}, {Friedlaender}, {Heyman}, {Hurlburt}, {Katz},
  {Kushner}, {Levay}, {Lindgren}, {Mathur}, {McFeaters}, {Mitchell}, {Rehse},
  {Schrijver}, {Springer}, {Stern}, {Tarbell}, {Wuelser}, {Wolfson}, {Yanari},
  {Bookbinder}, {Cheimets}, {Caldwell}, {Deluca}, {Gates}, {Golub}, {Park},
  {Podgorski}, {Bush}, {Scherrer}, {Gummin}, {Smith}, {Auker}, {Jerram},
  {Pool}, {Soufli}, {Windt}, {Beardsley}, {Clapp}, {Lang}, \&
  {Waltham}}]{2012SoPh..275...17L}
{Lemen}, J.~R., {et~al.} 2012, \solphys, 275, 17

\bibitem[{{Li} {et~al.}(2013){Li}, {Ning}, \& {Wang}}]{2013NewA...23...19L}
{Li}, D., {Ning}, Z.~J., \& {Wang}, J.~F. 2013, \na, 23, 19

\bibitem[{{Madjarska} {et~al.}(2003){Madjarska}, {Doyle}, {Teriaca}, \&
  {Banerjee}}]{2003A&A...398..775M}
{Madjarska}, M.~S., {Doyle}, J.~G., {Teriaca}, L., \& {Banerjee}, D. 2003,
  \aap, 398, 775

\bibitem[{{Mart{\'{\i}}nez-Sykora} {et~al.}(2011){Mart{\'{\i}}nez-Sykora}, {De
  Pontieu}, {Hansteen}, \& {McIntosh}}]{2011ApJ...732...84M}
{Mart{\'{\i}}nez-Sykora}, J., {De Pontieu}, B., {Hansteen}, V., \& {McIntosh},
  S.~W. 2011, \apj, 732, 84

\bibitem[{{McIntosh} {et~al.}(2014){McIntosh}, {Wang}, {Leamon}, \&
  {Scherrer}}]{2014ApJ...784L..32M}
{McIntosh}, S.~W., {Wang}, X., {Leamon}, R.~J., \& {Scherrer}, P.~H. 2014,
  \apjl, 784, L32

\bibitem[{{Nakagawa} \& {Raadu}(1972)}]{1972SoPh...25..127N}
{Nakagawa}, Y., \& {Raadu}, M.~A. 1972, \solphys, 25, 127

\bibitem[{{Nolte} {et~al.}(1979){Nolte}, {Solodyna}, \&
  {Gerassimenko}}]{1979SoPh...63..113N}
{Nolte}, J.~T., {Solodyna}, C.~V., \& {Gerassimenko}, M. 1979, \solphys, 63,
  113

\bibitem[{{Parnell} {et~al.}(1994){Parnell}, {Priest}, \&
  {Golub}}]{1994SoPh..151...57P}
{Parnell}, C.~E., {Priest}, E.~R., \& {Golub}, L. 1994, \solphys, 151, 57

\bibitem[{{P{\'e}rez-Su{\'a}rez} {et~al.}(2008){P{\'e}rez-Su{\'a}rez},
  {Maclean}, {Doyle}, \& {Madjarska}}]{2008A&A...492..575P}
{P{\'e}rez-Su{\'a}rez}, D., {Maclean}, R.~C., {Doyle}, J.~G., \& {Madjarska},
  M.~S. 2008, \aap, 492, 575

\bibitem[{{Pre{\'s} } \& {Phillips}(1999)}]{1999ApJ...510L..73P}
{Pre{\'s} }, P., \& {Phillips}, K.~H.~J. 1999, \apjl, 510, L73

\bibitem[{{Priest} {et~al.}(1994){Priest}, {Parnell}, \&
  {Martin}}]{1994ApJ...427..459P}
{Priest}, E.~R., {Parnell}, C.~E., \& {Martin}, S.~F. 1994, \apj, 427, 459

\bibitem[{{Schou} {et~al.}(2012){Schou}, {Scherrer}, {Bush}, {Wachter},
  {Couvidat}, {Rabello-Soares}, {Bogart}, {Hoeksema}, {Liu}, {Duvall}, {Akin},
  {Allard}, {Miles}, {Rairden}, {Shine}, {Tarbell}, {Title}, {Wolfson},
  {Elmore}, {Norton}, \& {Tomczyk}}]{2012SoPh..275..229S}
{Schou}, J., {et~al.} 2012, \solphys, 275, 229

\bibitem[{{Sheeley} \& {Golub}(1979)}]{1979SoPh...63..119S}
{Sheeley}, Jr., N.~R., \& {Golub}, L. 1979, \solphys, 63, 119

\bibitem[{{Srivastava} \& {Dwivedi}(2010)}]{2010MNRAS.405.2317S}
{Srivastava}, A.~K., \& {Dwivedi}, B.~N. 2010, \mnras, 405, 2317

\bibitem[{{Strong} {et~al.}(1992){Strong}, {Harvey}, {Hirayama}, {Nitta},
  {Shimizu}, \& {Tsuneta}}]{1992PASJ...44L.161S}
{Strong}, K.~T., {Harvey}, K., {Hirayama}, T., {Nitta}, N., {Shimizu}, T., \&
  {Tsuneta}, S. 1992, \pasj, 44, L161

\bibitem[{{Teriaca} {et~al.}(2004){Teriaca}, {Banerjee}, {Falchi}, {Doyle}, \&
  {Madjarska}}]{2004A&A...427.1065T}
{Teriaca}, L., {Banerjee}, D., {Falchi}, A., {Doyle}, J.~G., \& {Madjarska},
  M.~S. 2004, \aap, 427, 1065

\bibitem[{{Tian} {et~al.}(2008{\natexlab{a}}){Tian}, {Curdt}, {Marsch}, \&
  {He}}]{2008ApJ...681L.121T}
{Tian}, H., {Curdt}, W., {Marsch}, E., \& {He}, J. 2008{\natexlab{a}}, \apjl,
  681, L121

\bibitem[{{Tian} {et~al.}(2011{\natexlab{a}}){Tian}, {McIntosh}, \& {De
  Pontieu}}]{2011ApJ...727L..37T}
{Tian}, H., {McIntosh}, S.~W., \& {De Pontieu}, B. 2011{\natexlab{a}}, \apjl,
  727, L37

\bibitem[{{Tian} {et~al.}(2011{\natexlab{b}}){Tian}, {McIntosh}, {De Pontieu},
  {Mart{\'{\i}}nez-Sykora}, {Sechler}, \& {Wang}}]{2011ApJ...738...18T}
{Tian}, H., {McIntosh}, S.~W., {De Pontieu}, B., {Mart{\'{\i}}nez-Sykora}, J.,
  {Sechler}, M., \& {Wang}, X. 2011{\natexlab{b}}, \apj, 738, 18

\bibitem[{{Tian} {et~al.}(2012){Tian}, {McIntosh}, {Wang}, {Ofman}, {De
  Pontieu}, {Innes}, \& {Peter}}]{2012ApJ...759..144T}
{Tian}, H., {McIntosh}, S.~W., {Wang}, T., {Ofman}, L., {De Pontieu}, B.,
  {Innes}, D.~E., \& {Peter}, H. 2012, \apj, 759, 144

\bibitem[{{Tian} {et~al.}(2008{\natexlab{b}}){Tian}, {Xia}, \&
  {Li}}]{2008A&A...489..741T}
{Tian}, H., {Xia}, L.-D., \& {Li}, S. 2008{\natexlab{b}}, \aap, 489, 741

\bibitem[{{Tian} {et~al.}(2014){Tian}, {DeLuca}, {Cranmer}, {De Pontieu},
  {Peter}, {Mart{\'{\i}}nez-Sykora}, {Golub}, {McKillop}, {Reeves}, {Miralles},
  {McCauley}, {Saar}, {Testa}, {Weber}, {Murphy}, {Lemen}, {Title}, {Boerner},
  {Hurlburt}, {Tarbell}, {Wuelser}, {Kleint}, {Kankelborg}, {Jaeggli},
  {Carlsson}, {Hansteen}, \& {McIntosh}}]{2014Sci...346A.315T}
{Tian}, H., {et~al.} 2014, Science, 346, 1255711

\bibitem[{{Torrence} \& {Compo}(1998)}]{1998BAMS...79...61T}
{Torrence}, C., \& {Compo}, G.~P. 1998, Bulletin of the American Meteorological
  Society, 79, 61

\bibitem[{{Ugarte-Urra} {et~al.}(2004){Ugarte-Urra}, {Doyle}, {Madjarska}, \&
  {O'Shea}}]{2004A&A...418..313U}
{Ugarte-Urra}, I., {Doyle}, J.~G., {Madjarska}, M.~S., \& {O'Shea}, E. 2004,
  \aap, 418, 313

\bibitem[{{Vaiana} {et~al.}(1973){Vaiana}, {Davis}, {Giacconi}, {Krieger},
  {Silk}, {Timothy}, \& {Zombeck}}]{1973ApJ...185L..47V}
{Vaiana}, G.~S., {Davis}, J.~M., {Giacconi}, R., {Krieger}, A.~S., {Silk},
  J.~K., {Timothy}, A.~F., \& {Zombeck}, M. 1973, \apjl, 185, L47

\bibitem[{{Wang} \& {Shi}(1993)}]{1993SoPh..143..119W}
{Wang}, J., \& {Shi}, Z. 1993, \solphys, 143, 119

\bibitem[{{Zhang} {et~al.}(2001){Zhang}, {Kundu}, \&
  {White}}]{2001SoPh..198..347Z}
{Zhang}, J., {Kundu}, M.~R., \& {White}, S.~M. 2001, \solphys, 198, 347

\bibitem[{{Zhang} {et~al.}(2014){Zhang}, {Chen}, {Ding}, \&
  {Ji}}]{2014A&A...568A..30Z}
{Zhang}, Q.~M., {Chen}, P.~F., {Ding}, M.~D., \& {Ji}, H.~S. 2014, \aap, 568,
  A30

\bibitem[{{Zhang} {et~al.}(2012){Zhang}, {Chen}, {Guo}, {Fang}, \&
  {Ding}}]{2012ApJ...746...19Z}
{Zhang}, Q.~M., {Chen}, P.~F., {Guo}, Y., {Fang}, C., \& {Ding}, M.~D. 2012,
  \apj, 746, 19

\end{thebibliography}

\end{document}